\documentclass[12pt]{iopart}
\usepackage{graphicx}
\usepackage{cite}
\usepackage{hyperref}
\begin{document}

\title[Limitations of polarisation optics for the polarisation Sagnac speedmeter]{Experimental investigation of the limitations of polarisation optics for future gravitational wave detectors based on the polarisation Sagnac speedmeter}

\author{{A.~P.~Spencer$^{1}$}, {B.~W.~Barr$^1$}, {A.~S.~Bell$^1$}, {J.~Briggs$^1$}, {P.~Dupej$^1$}, {S.~H.~Huttner$^1$}, {B.~Sorazu$^1$}, {J.~Wright$^1$} and {K.~A.~Strain$^1$}}

\address{$^1$Institute for Gravitational Research, University of Glasgow, Glasgow UK}
\ead{andrew.spencer@glasgow.ac.uk}
\vspace{10pt}
\begin{indented}
\item[]July 2021
\end{indented}

\begin{abstract}
The polarisation Sagnac speedmeter interferometer has the potential to replace the Michelson interferometer as the instrumental basis for future generations of ground-based gravitational wave detectors. The quantum noise benefit of this speedmeter is dependent on high-quality polarisation optics, the polarisation beam-splitter (PBS) and quarter-waveplate (QWP) optics that are key to this detector configuration and careful consideration of the effect of birefringence in the arm cavities of the interferometer. A PBS with an extinction ratio of better than 4000 in transmission and 700 in reflection for a $41^{\circ}$ angle of incidence was characterised along with a QWP of birefringence of $\case{\lambda}{4} + \case{\lambda}{324}$. The cavity mirror optics of a 10\,m prototype polarisation Sagnac speedmeter were measured to have birefringence in the range $1\times10^{-3}$ to $2\times10^{-5}$\,radians. This level of birefringence, along with the QWP imperfections, can be canceled out by careful adjustment of the QWP angle, to the extent that the extinction ratio of the PBS is the leading limitation for the polarisation Sagnac speedmeter in terms of polarisation effects.

\end{abstract}
\section{Introduction}
Operational gravitational wave observatories, including the two LIGO  detectors in the USA \cite{aLIGO} and the Virgo  detector in Italy \cite{aVirgo} have been highly successful at measuring gravitational waves emitted by the coalescence of black-hole and neutron star binary systems \cite{{O1O2cat1},{GWTC2}}. These detectors are based on Michelson interferometer configurations with suspended test masses where information about the relative position of the test masses is recovered from the phase of the laser light at the interferometer readout after the light has been modulated by interaction with a gravitational wave \cite{InterferometryLivRev}.

As improvements are made to these detectors to decrease the contribution of classical  noise  \cite{{Buikema20},{Acernese19}}, the detectors become quantum noise limited. Sensitivity in the detectors upper frequency band, where quantum shot noise is the dominant effect,  can be improved by increasing the laser power \cite{Bode20} and using squeezed light injection \cite{{Tse19},{Mehmet20}}. This comes at the penalty of increased quantum radiation pressure noise at lower frequencies, although this currently remains below the classical noise level in this band.

The dissimilar relationship between optical power and the two quantum noise terms, along with the frequency dependence of both, results in an optimal power at each frequency  that results in the best sensitivity of the instrument that is known as the \textit{standard quantum limit} of a detector \cite{Corbitt04}.

One way to improve on the sensitivity limitation imposed by the standard quantum limit of a Michelson interferometer is to consider a different interferometer topology such as the speedmeter \cite{FirstSpeedmeter} with an inherently lower standard quantum limit.

In a speedmeter interferometer the phase of the light at the readout contains information about the relative velocity (hence \textit{speed}meter) of the test masses by interferometric cancellation of positional information. The quantum radiation pressure noise is thus cancelled, resulting in lower quantum noise ideally limited only by the quantum shot noise \cite{QMTLivRev}.

There are multiple topologies of speedmeter that are viable for future gravitational wave detectors such as the Einstein Telescope \cite{ET10}. In this paper we consider the polarisation Sagnac speedmeter \cite{DanilishinPolSM} which unlike the ring-cavity Sagnac \cite{{ChenRing}} has linear arm cavities, and does not require additional cavities as in the sloshing speedmeter designs \cite{{PurdueChenSlosh},{Huttner},{DanilishinPolCirc},{EPR}}. This make the polarisation Sagnac speedmeter attractive to consider as it is both conceptually straightforward and employs, for the most part, components and techniques that are already familiar in Michelson-based detectors and has the advantage of linear arms that are compatible with existing and planned infrastructures.

\begin{figure}
\centering
\includegraphics[width=0.75\textwidth]{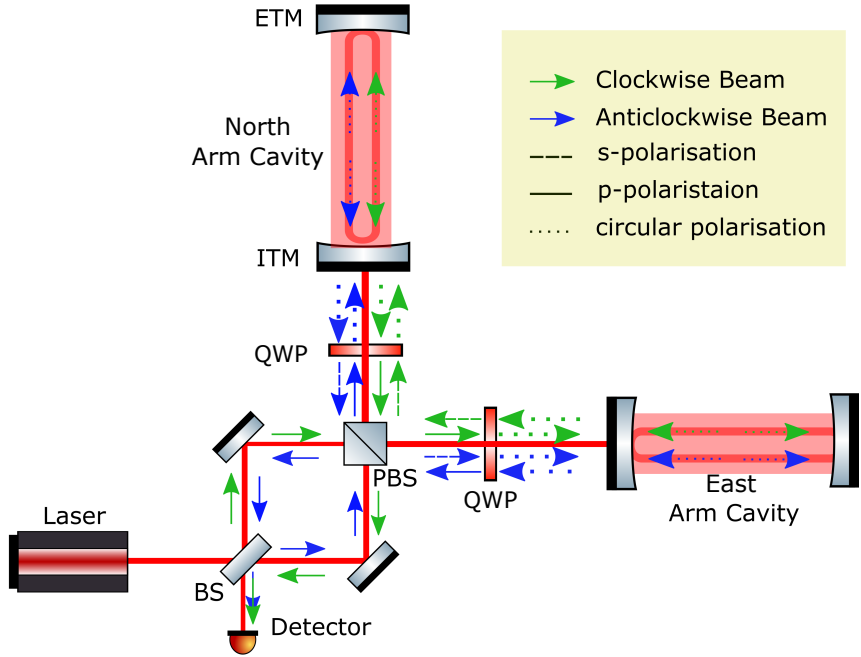}
\caption{The polarisation Sagnac speedmeter as a gravitational wave detector interferometer. Laser light of linear polarisation illuminating the interferometer from the west is split into two equal counter-propagating beams by the beam-splitter (BS). The clockwise beam (green) is first transmitted by the polarising beam-splitter (PBS) and resonates in the east arm cavity acquiring positional information about this pair of test masses. Returning to the PBS, as light of the orthogonal linear polarisation having double-passed the quarter-wave plate (QWP), the beam is reflected towards the north arm cavity interacting with the second test mass pair after a time delay due to the light travel time. Upon a third interaction with the PBS the light is of the original linear polarisation, due to a second double-pass of the northern QWP, so is transmitted back to the BS where it interferes with the counter-clockwise beam (blue) which has followed the reverse path. The result of the interference is a beam of light, phase modulated with information about the relative differential velocity of the two test mass cavities.}
\label{fig:PolSagnac}
\end{figure}

%%%%%%%%%%%%%%%%%%%%%%%%%%%%%%%%%%%%%%%%%%%%%%%%%%%%%%%%%%%%%%%%%%%%%%%%%%%%%%%%%%%%%%%%%%%%%%%%%%%%%%%%%%%%%%%%%%%%%%%%%%%%%%
%             Section 2: Speedmeter Limitations
%%%%%%%%%%%%%%%%%%%%%%%%%%%%%%%%%%%%%%%%%%%%%%%%%%%%%%%%%%%%%%%%%%%%%%%%%%%%%%%%%%%%%%%%%%%%%%%%%%%%%%%%%%%%%%%%%%%%%%%%%%%%%%
\section{The Polarisation Sagnac Speedmeter}
\begin{figure}
    \centering
    \includegraphics[width=0.95\textwidth]{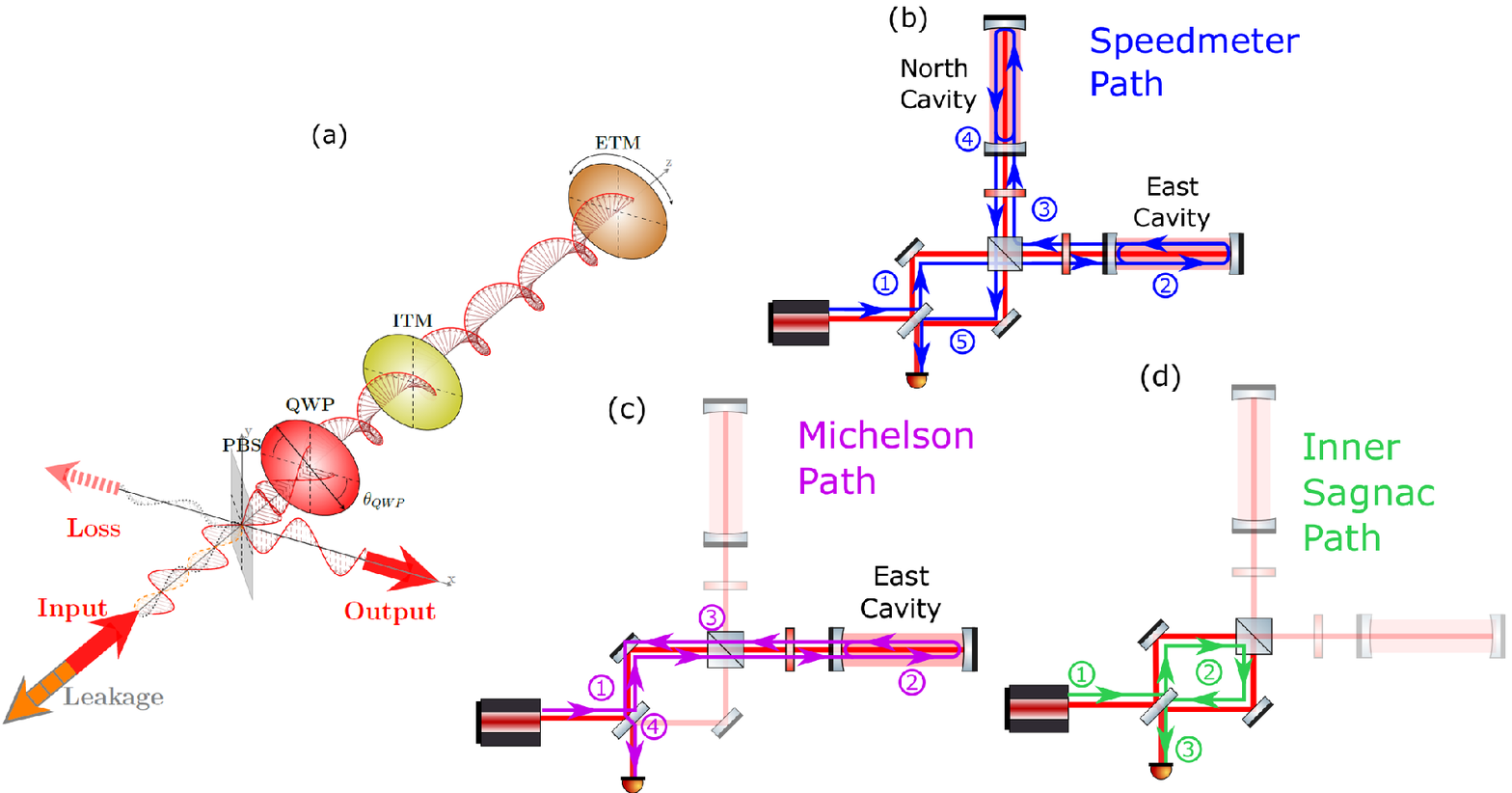}
    \caption{(a) Single arm of the polarisation Sagnac speedmeter showing the input, output and leakage of light within the interferometer due to imperfect polarisation optics. (b-d) Possible optical paths of the clockwise propagating beam though the interferometer from input to output. Light following the desired speedmeter path (b) will be resonant in both arm cavities. Light following the Michelson path (c) due to a combination of imperfect separation of polarisation at each PBS port and unwanted birefringence will only sense one arm of the interferometer. Also due to the imperfect separation of polarisation at the PBS some light will not enter either cavity and will circulate between the steering mirrors in the so-called inner-Sagnac path (d).}
    \label{fig:LinCav}
\end{figure}

In a polarisation Sagnac speedmeter (Figure \ref{fig:PolSagnac}) the polarising optics are configured to produce two counter-propagating beam paths such that light probes each arm cavity sequentially. This introduces a delay between interactions with the two sets of test masses that results in a phase modulation that is dependent on the relative velocity of the test mass pairs; in the ideal case with no dependence on the relative positions of the test masses.

Imperfections in the polarisation optics will degrade the performance of the Sagnac speedmeter by reintroducing positional phase dependence (or Michelson signal) into the output field and resulting in imperfect cancellation of quantum radiation pressure noise \cite{AsymmerticSSM}.

The possible paths that light can take to introduce additional quantum noise to the output due to polarisation imperfections are shown in Figure \ref{fig:LinCav}. We use the relative amount of position-dependent signal at the readout port (or the amount of Michelson-to-Speedmeter signal) as a metric to judge the adverse effect of imperfections in the polarisation optics.  In this paper we only consider polarisation effects and no other optical loss mechanisms.

%%%%%%%%%%%%%%%%%%%%%%%%%%%%%%%%%%%%%%%%%%%%%%%%%%%%%%%%%%%%%%%%%%%%%%%%%%%%%%%%%%%%%%%%%%%%%%%%%%%%%%%%%%%%%%%%%%%%%%%%%%%%%%
%             Section 3: PBS
%%%%%%%%%%%%%%%%%%%%%%%%%%%%%%%%%%%%%%%%%%%%%%%%%%%%%%%%%%%%%%%%%%%%%%%%%%%%%%%%%%%%%%%%%%%%%%%%%%%%%%%%%%%%%%%%%%%%%%%%%%%%%%
\section{Characterisation of a Polarisation Beam-Splitter}\label{sec:PBS}

\begin{figure}
    \centering
    \includegraphics[width=0.85\textwidth]{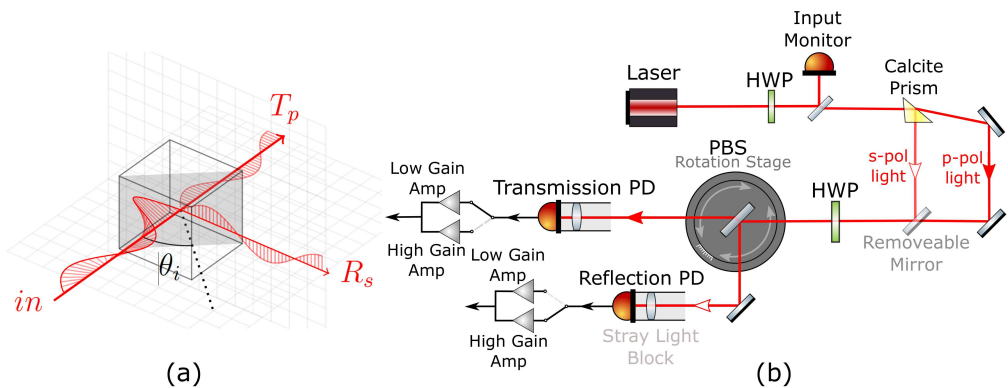}
    \caption{(a) Polarising beam-splitter: the input field of arbitrary polarisation is split into two orthogonally polarised beams in reflection and transmission of the optical surface (gray shaded area). (b) Experimental set-up for the characterisation of the extinction ratio as a function of the angle of incidence of a thin-film polariser (PBS) on a Vernier-scale rotation stage. The input light was prepared in a linear polarisation aligned to the PBS with a calcite prism and a half-waveplate. The set-up could be switched between a p- or s-polarised input beam with use of a removable mirror and beam blocks. The reflected and transmitted beams are recorded on separate photodiodes with a third as an input monitor.}
    \label{fig:PBS_EXP}
\end{figure}

The experimental setup to measure the polarisation extinction ratio in transmission and reflection of the thin-film PBS is shown in Figure \ref{fig:PBS_EXP}\,(b). The optics under investigation were Layertec \cite{Layertech} thin film polarisers with a specified polarisation extinction ratio in transmission ($\textrm{PER}_T$) of 2500 for 1550nm light at an angle of incidence of $45^{\circ}$. Thin film polarisers are chosen as they are compatible with the suspension systems required of the core optics of a gravitational wave detector and can be more economically scaled to larger sizes in comparison to prism polarisers. 

The light is prepared in a linear polarisation using a calcite prism of extremely high polarisation extinction ratio \cite{Calcite} and a half waveplate (HWP) to align the polarisation of the light to the PBS. In this way the extinction ratio could be maximised such that the best performance of the optic could be investigated without limits imposed by misalignment between the input polarisation and the PBS being characterised.

The reflected and transmitted beams are recorded on separate photodiodes with calibrated high and low gain readout electronics that were designed to probe high extinction ratios by being able to measure light power of a few milli-watts or a few micro-watts respectively. A third photodiode was used to monitor the optical power at the input of the system. The input monitor was used as a calibrated readout of the power in-front of the PBS during the measurements to mitigate possible measurement errors introduced by changing input power.

\begin{figure}
    \centering
    \includegraphics[width=0.7\textwidth]{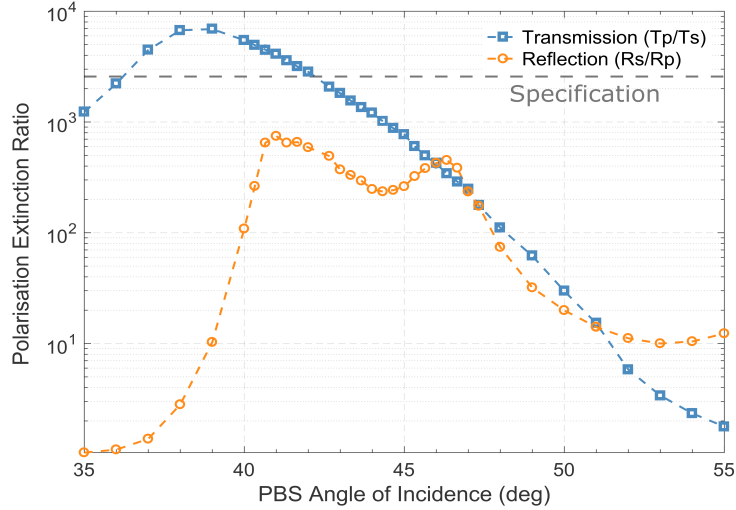}
    \caption{Measured polarisation extinction ratio in transmission (blue) and reflection (orange) of the PBS as a function of angle of incidence compared in the specification of 2500.}
    \label{fig:PBS_PER}
\end{figure}

The extinction ratio in transmission ($\textrm{PER}_{T} = \case{T_p}{T_s}$) and reflection ($\textrm{PER}_{R} = \case{R_s}{R_p}$) as a function of angle of incidence on the PBS was measured and the results are shown in Figure \ref{fig:PBS_PER}. At an angle of incidence of $45^{\circ}$ the extinction ratio in transmission was measured to be $770\pm6.5$, and $262\pm1.8$ in reflection. The optimal angle of incidence for concurrent transmission and reflection extinction was found to be around $41^{\circ}$, where the transmission extinction ratio was measured to be better than 4000, and 700 in reflection compared to the specified transmission extinction ratio of 2500.
%%%%%%%%%%%%%%%%%%%%%%%%%%%%%%%%%%%%%%%%%%%%%%%%%%%%%%%%%%%%%%%%%%%%%%%%%%%%%%%%%%%%%%%%%%%%%%%%%%%%%%%%%%%%%%%%%%%%%%%%%%%%%%
%             Section 4: QWP
%%%%%%%%%%%%%%%%%%%%%%%%%%%%%%%%%%%%%%%%%%%%%%%%%%%%%%%%%%%%%%%%%%%%%%%%%%%%%%%%%%%%%%%%%%%%%%%%%%%%%%%%%%%%%%%%%%%%%%%%%%%%%%
\section{Characterisation of a Quarter-Waveplate Optic}\label{sec:QWP}

Imperfections in the manufacture of a quarter-waveplate (QWP) (thickness, surface uniformity, etc) will result in imperfect polarisation transformation and allow Michelson signal leakage in the polarisation Sagnac speedmeter system.

The waveplates purchased for the polarisation Sagnac speedmeter experiment were optically contacted, zero-order waveplates from Union Optic \cite{UOptic} with a specified birefringence and tolerance of $\case{\lambda}{4} \pm \case{\lambda}{300}$ for 1550\,nm light.

\begin{figure}
    \centering
    \includegraphics[width=0.85\textwidth]{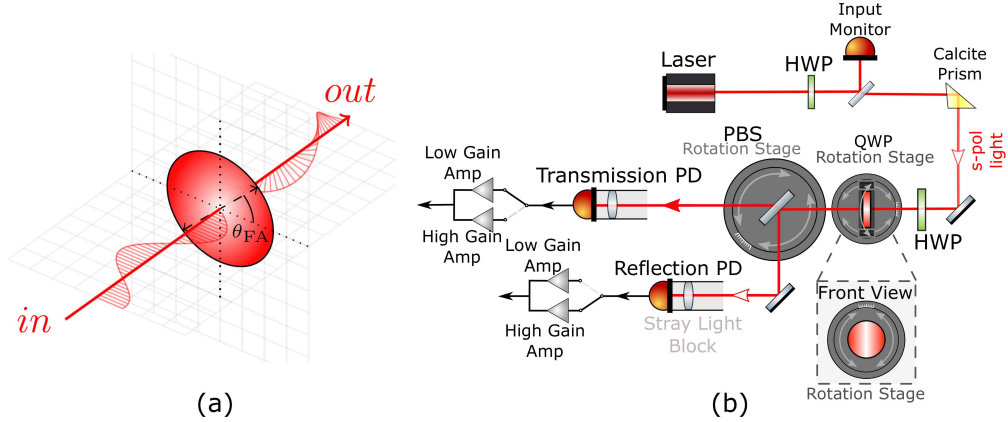}
    \caption{ (a) Basic principle of a quarter waveplate. The input field of linear polarisation is rotated into a circular polarisation by the QWP with the fast axis orientated at $45^{\circ}$ to the input polarisation. (b) Experimental setup to measure the birefringence of a QWP as a function of fast axis orientation. The QWP is mounted in a rotating mount on a rotation stage so both the fast axis angle and the angle of incidence can be adjusted for optimal performance. The input light is prepared in a linear polarisation with a calcite prism and a half-waveplate (HWP) and the input power is monitored with a photodiode (PD) pick-off before the calcite prism. The polarisation state of light in transmission of the QWP is analysed with a PBS with PD readout in each output port.}
    \label{fig:FigQWP}
\end{figure}

The experimental setup to measure the birefringence of a quarter-waveplate is shown in Figure \ref{fig:FigQWP}\,(b). The birefringence of the QWP is measured as the induced change in polarisation as a function of the rotation of the QWP fast axis.

In this setup an ideal quarter-waveplate at an angle of $45^{\circ}$ relative to the linearly polarised light should produce the greatest rotation of the polarisation of the laser light. The combination of the calcite prism and HWP was set to produce s-polarised light (in the frame of the QWP) and in the case of an ideal QWP would result in an equal amount of optical power reflected and transmitted by the PBS. The extinction ratio of the PBS as characterised in Section \ref{sec:PBS} is taken into account when analysing the resulting polarisation of light in transmission of the QWP.

\begin{figure}
    \centering
    \includegraphics[width=0.7\textwidth]{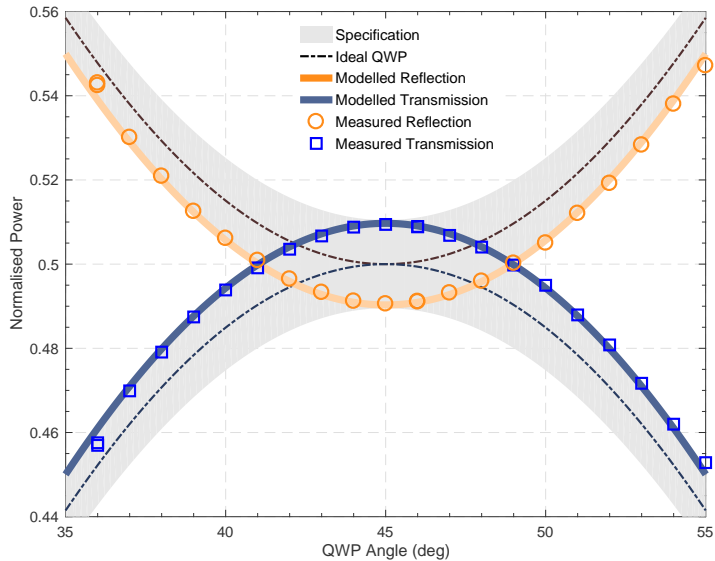}
    \caption{Result of the characterisation experiment for the QWP optic. The measured optical power in reflection (orange) and transmission (blue) of the PBS was used to characterise the birefringence of the QWP optic compared to the specification (grey region). The predicted result for an ideal QWP is also shown (dashed lines). The modelled result with a fitted value of QWP birefringence is also shown fitted to the data.}
    \label{fig:QWP_Char}
\end{figure}

Plotting the light transmitted and reflected by the PBS as a function of the angle of the QWP fast axis, the results can be compared to a model of the setup using Jones calculus \cite{PolShurcliff} to describe the polarisation optics and the polarisation vector of the laser light (see Figure \ref{fig:QWP_Char}). Fitting this model to the data with the QWP birefringence as a free parameter gives a birefringence of $1.590 \pm 0.002$\,radians. This result can be expressed as $\case{\lambda}{4} + \case{\lambda}{324}$ which is within the manufactures specification of the optic having a birefringence tolerance of $\case{\lambda}{4} + \case{\lambda}{300}$.

%%%%%%%%%%%%%%%%%%%%%%%%%%%%%%%%%%%%%%%%%%%%%%%%%%%%%%%%%%%%%%%%%%%%%%%%%%%%%%%%%%%%%%%%%%%%%%%%%%%%%%%%%%%%%%%%%%%%%%%%%%%%%%
%             Section 5: Cavity Birefringence
%%%%%%%%%%%%%%%%%%%%%%%%%%%%%%%%%%%%%%%%%%%%%%%%%%%%%%%%%%%%%%%%%%%%%%%%%%%%%%%%%%%%%%%%%%%%%%%%%%%%%%%%%%%%%%%%%%%%%%%%%%%%%%
\section{Investigation of Cavity Birefringence}\label{sec:Cav}
Birefringence in multi-layer amorphous mirror coatings, such as those used for gravitational wave detectors, has previously been reported on. \cite{{BirefringenceMoriwaki},{BirefringenceBrandi},{BirefringenceJacob}}, with the level of birefringence measured ranging between $10^{-4}$ to $10^{-7}$ radians per reflection \cite{BirefringenceReview}. Higher birefringence can be expected in crystalline coatings which are also of interest for future gravitational wave detectors \cite{Cole16}.

In multi-layer coatings, stress-induced birefringence can result from the coating process after the optical layers have been deposited due to the difference in thermal properties of the coating and the substrate material. 

Birefringence is present in the mirrors for current gravitational wave detectors but as the detection method does not directly rely on the polarisation state of light in the interferometer the impact is less significant and has been shown to be a minor source of loss in such an interferometer \cite{Winkler}.

In a polarisation Sagnac speedmeter the polarisation state of the light in the detector arms is critical and even low-levels of birefringence in the cavity mirrors can have a significant effect on the the overall performance as the effect of mirror coating birefringence scales with the finesse of the arm cavity.

\begin{figure}
    \centering
    \includegraphics[width=0.75\textwidth]{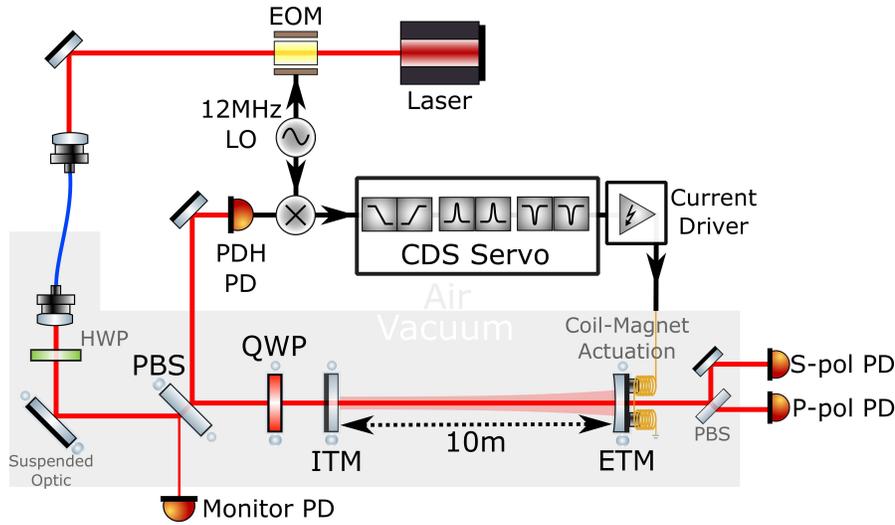}
    \caption{Experimental layout to measure the polarisation ellipticity induced by birefringence effects in the mirrors forming a 10\,m long Fabry-Perot cavity. The pre-stabilised laser is locked to the cavity length using the Pound-Drever-Hall (PDH) technique to ensure the laser light is resonant in the arm cavity. The PDH error signal is shaped by a digital servo implemented in a CDS unit\cite{CDS} that consists of gain shaping low-pass transitional integrator and differentiator filters to lock the cavity, and resonant gain and notch filters to compensate for the pendulum and mechanical modes of the suspended optics. The length of the cavity is controlled by applying a feedback signal to a coil-magnet actuator on the suspended ETM mirror to lock the cavity to the laser with a unity gain frequency of 600 Hz. The transmitted light is analysed using a PBS with readout photodiode in reflection and transmission. This PBS was calibrated with the ETM mirror removed to measure the polarisation of the light entering the cavity with no cavity birefringence effects. More details about the suspended 10\,m cavity infrastructure can be found in \cite{SpencerPhD}.}
    \label{fig:Cav_Exp}
\end{figure}

An experiment was devised to measure the birefringence of the cavity mirrors for the prototype polarisation Sagnac speedmeter. This involves measuring the polarisation state of the light transmitted by a suspended 10\,m Fabry-Perot cavity to infer the birefringence effect of the cavity mirrors. The layout for this experiment is shown in Figure \ref{fig:Cav_Exp}.

The polarisation state of the light in transmission of the cavity on resonance is analysed with a PBS with two photodiodes in reflection and transmission. This readout system was calibrated by first measuring the polarisation of the light that enters the cavity by removing the ETM from the optical path so that no cavity is present and the beam transmitted by the ITM could be analysed. The ellipticity induced by the cavity can be measured by comparing this no-cavity result to the case where the cavity is held on resonance with the input laser light.

This ellipticity measurement was performed for six different relative orientations of the cavity mirrors; achieved by rotating the ETM optic in its suspension. Rotating the ETM relative to the ITM changes the orientation of the fast axis of the ETM and therefore the combined birefringence of the cavity. The resulting change in the ellipticity of the transmitted cavity light would be expected to follow a sinusoidal relationship with ETM angle as the mirror is in effect a low birefringence waveplate.

\begin{figure}
    \centering
    \includegraphics[width=0.9\textwidth]{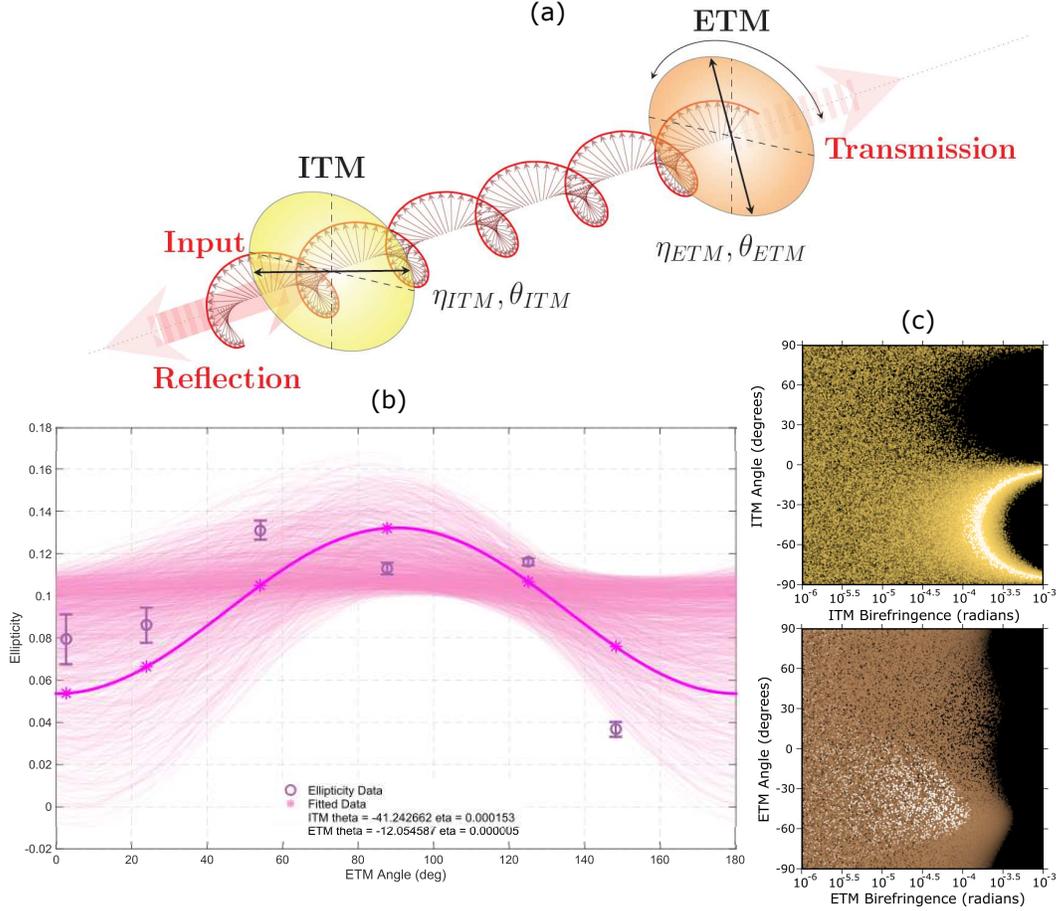}
    \caption{Results of the experiment to measure cavity birefringence. In a Fabry-Perot cavity (a) where the ITM and ETM mirrors are considered to be birefringent each mirror can be considered as a weak waveplate with birefringence ($\eta$) and fast axis orientated at angle ($\theta$) to the horizontal plane. The cavity is illuminated with circularly polarised light which becomes elliptically polarised while resonating in the cavity. This ellipticity of the transmitted light was measured (b) for six different orientations of the ETM fast axis ($\theta_{\textrm{ETM}}$) relative to the ITM. A model was used to preform a parameter estimation fit to the data (purple) and each pink trace in (b) is a possible fit that corresponds to the white point in (c) of best-guess parameters for the two birefringent mirrors.}
    \label{fig:Ellip}
\end{figure}

In order to interpret the results a model of the experiment was built, based on Jones calculus. This model is the linear combination of Jones matrices \cite{PolShurcliff} that represent each polarising optical element as arranged in Figure \ref{fig:Cav_Exp}. The birefringence of each cavity mirror is included in the model as a weak waveplate and the combination forms a Fabry-Perot cavity that induces a degree of ellipticity to the circulating light. The model was used to generate ellipticity curves that were compared to the experimental data shown in Figure \ref{fig:Ellip}\,(b). More detail on the modelling procedure can be found in \cite{SpencerPhD}.

In the model the reflection and transmission coefficients of the  ITM and ETM are known parameters. While the birefringence amplitude and fast axis angles of both mirrors are unknown parameters. A Monte Carlo method was used to generate multiple models of the experiment with combinations of the unknown parameters selected from a defined parameter space. The parameter space limits for the mirror birefringence were set based on \cite{BirefringenceReview} and a half rotation of the fast axis of each mirror was considered due to symmetry. Comparing the fit of each test to the data with a quality factor, it was possible to produce a parameter estimation result shown in Figure \ref{fig:Ellip}\,(c) where each point in this plot represents a single model in the parameter space. A few thousand simulations were preformed to ensure the full parameter space was considered. The colour shading for each point represents the quality of fit with white being the best fit.

From this result bounds can be placed on the mirror's birefringent properties. The measured data was found to be consistent with an ITM of birefringence in the range from  $1.3\times10^{-4}$\,radians to $1.0\times10^{-3}$\,radians, and an ETM of birefringence in the range from $2.0\times10^{-5}$\,radians to $9.1\times10^{-5}$\,radians.

%%%%%%%%%%%%%%%%%%%%%%%%%%%%%%%%%%%%%%%%%%%%%%%%%%%%%%%%%%%%%%%%%%%%%%%%%%%%%%%%%%%%%%%%%%%%%%%%%%%%%%%%%%%%%%%%%%%%%%%%%%%%%%
%             Section 6: Speedmeter Consequences
%%%%%%%%%%%%%%%%%%%%%%%%%%%%%%%%%%%%%%%%%%%%%%%%%%%%%%%%%%%%%%%%%%%%%%%%%%%%%%%%%%%%%%%%%%%%%%%%%%%%%%%%%%%%%%%%%%%%%%%%%%%%%%
\section{Consequences for the Polarisation Sagnac Speedmeter}\label{sec:Comb}

\begin{figure}
   \centering
    \includegraphics[width=0.75\textwidth]{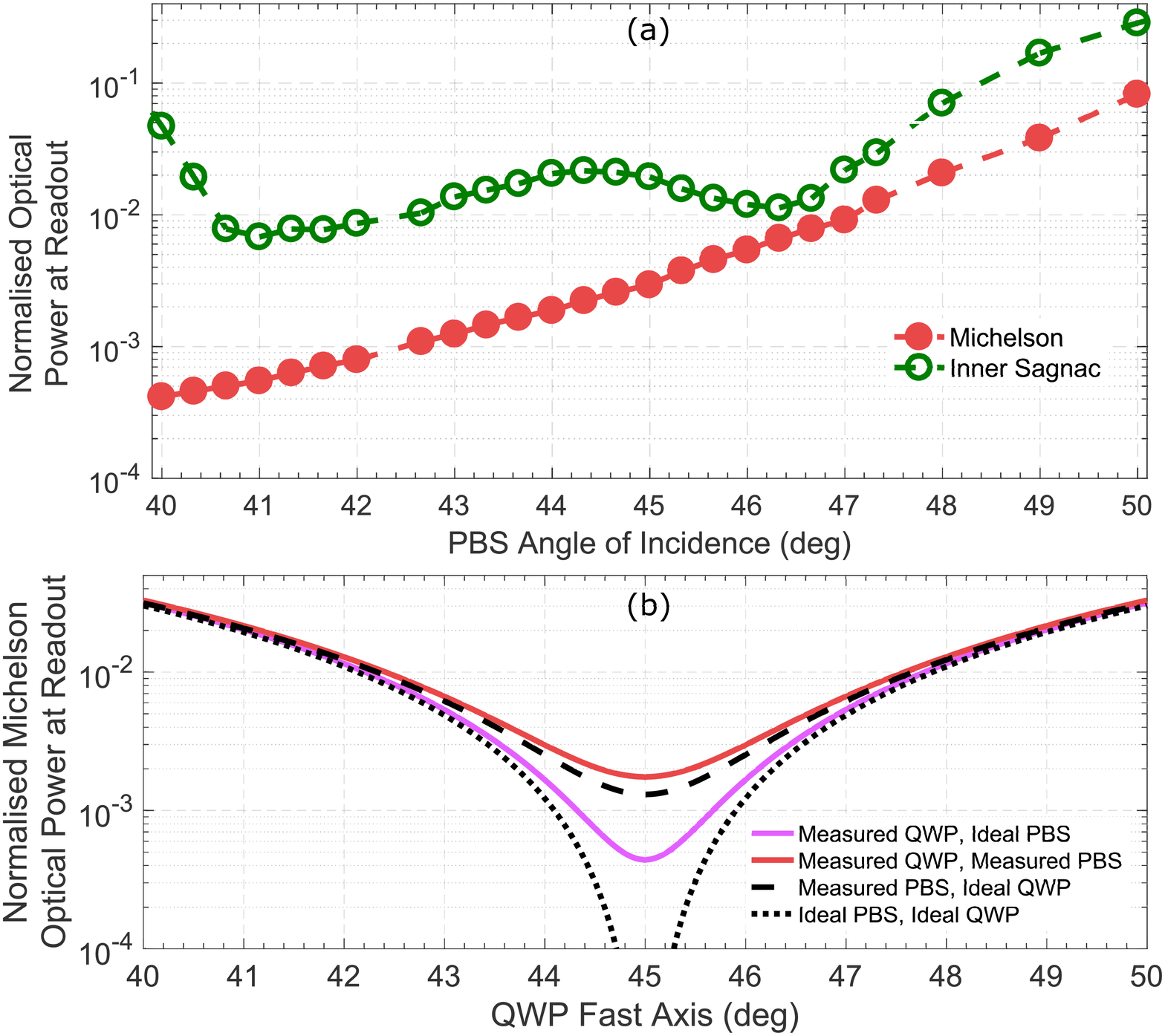}
    \caption{Effect of imperfect PBS and QWP optics on the amount of unwanted light at the readout of a polarisation Sagnac speedmeter. The results are normalised to the total optical power at the readout. For the PBS (a) the result is shown for three paths through a full polarisation Sagnac speedmeter as a function of the PBS angle of incidence. For the QWP (b) the fraction of optical power at the output of a single arm of the speedmeter that is Michelson leakage is shown as a function of QWP fast axis angle.}
    \label{fig:PBSQWP_Mod}
\end{figure}

A model of the polarisation Sagnac speedmeter was constructed using Jones calculus to model the polarisation state of light following the possible paths though the interferometer. This model allows us to determine the amount of light reaching the readout of the interferometer that follows the Michelson path and will therefore degrade the sensitivity of the instrument as a speedmeter. 

\subsection{Impact of PBS imperfection}
The modelled result for the PBS characterised in Section \ref{sec:PBS} is given in Figure \ref{fig:PBSQWP_Mod}\,(a) where the amount of optical power at the readout that has followed undesirable paths is shown.

Light following the \textit{Inner Sagnac} path contributes the most at the output. This can be considered a source of optical loss in the instrument as the light does not couple into either cavity therefore does not contribute to the detector sensitivity. It is the \textit{Michelson} path which will most severely impact the detector sensitivity by reintroducing positional information at the readout.

The result shows that at a $45^{\circ}$ angle of incidence $0.295\%$ of the light at the readout will have followed the Michelson path and this Michelson signal decreases below $0.1\%$ for an angle of incidence $<43^{\circ}$.

\subsection{Impact of QWP imperfection}
The amount of Michelson leakage from one arm of the speedmeter due to a QWP with a birefringence of $\case{\lambda}{4} + \case{\lambda}{324}$ is shown in Figure \ref{fig:PBSQWP_Mod}\,(b) as a function of the QWP fast axis rotation. This is shown for the case of an ideal PBS and for the PBS measured extinction ratio at the optimal angle of incidence.

In a full speedmeter interferometer with two arm QWPs of the same birefringence $0.088\%$ of the light at the readout will have followed a Michelson path through the interferometer this increases to $0.348\%$ in combination with the imperfect PBS.

\begin{figure}
   \centering
    \includegraphics[width=0.95\textwidth]{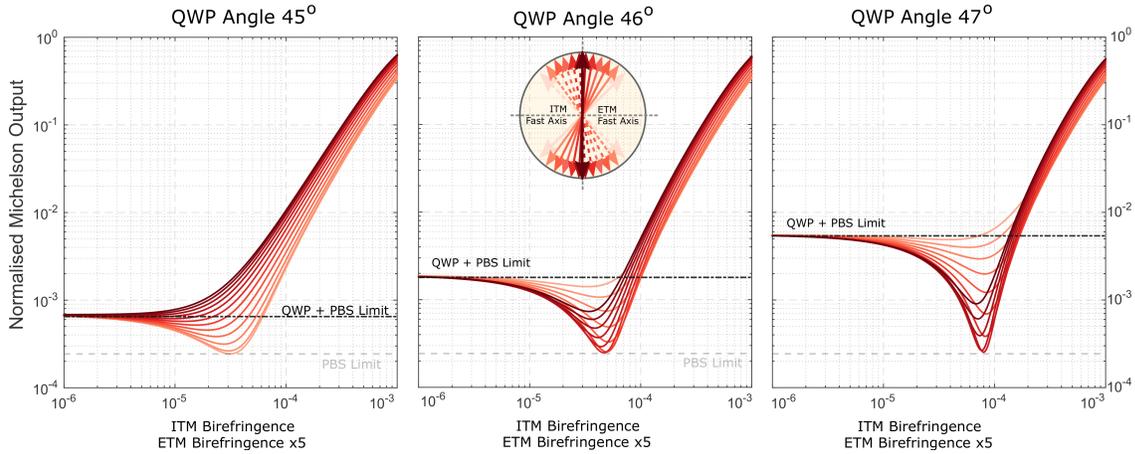}
    \caption{Modelling the combined effect of imperfect PBS and QWP optics in one arm of a polarisation Sagnac speedmeter with birefringent cavity mirrors. The level of Michelson leakage signal (normalised to optical power at the output) is plotted against the birefringence of the cavity mirrors for the case that the ETM is five times more birefringent than the ITM based on the limits obtained in Section \ref{sec:Cav}. The three plots correspond to three different orientations of the QWP fast axis relative to the cavity mirrors. The colour of each trace relates to the relative orientation of the two cavity mirrors as shown in the insert in the middle plot.}
    \label{fig:Cav_Mod}
\end{figure}

\subsection{Impact of cavity birefringence}
In the three sub-plots of Figure \ref{fig:Cav_Mod} the limitation on the amount of light following the Michelson path due solely to the imperfect PBS is shown as a grey dashed line along with the limit due to the combination of imperfect PBS and QWP as a black dashed line. For low levels of cavity birefringence ($\eta_{ITM} < 10^{-5}$ radians) it is the combination of PBS and QWP imperfections that limit the performance of the speedmeter. Cavity birefringence becomes the limiting factor for high levels of birefringence ( $\eta_{ITM} > 10^{-4}$ radians). In the region between these two limits there is a cancellation effect between the birefringent cavity and the imperfect QWP that depends on the relative orientation of these three optics. This can be seen in Figure \ref{fig:Cav_Mod} where for certain orientations of the QWP and cavity optics the level of Michelson signal drops to the limit imposed by the imperfect PBS optic. This manipulation of `imperfect' birefringence of the QWP via rotation of the optic can cancel a range of cavity birefringence (where the range is determined by the level of birefringent imperfection in the QWP) and mean the polarisation Sagnac is limited only by the PBS extinction ratio.

%%%%%%%%%%%%%%%%%%%%%%%%%%%%%%%%%%%%%%%%%%%%%%%%%%%%%%%%%%%%%%%%%%%%%%%%%%%%%%%%%%%%%%%%%%%%%%%%%%%%%%%%%%%%%%%%%%%%%%%%%%%%%%
%             Section 7: Conclusions
%%%%%%%%%%%%%%%%%%%%%%%%%%%%%%%%%%%%%%%%%%%%%%%%%%%%%%%%%%%%%%%%%%%%%%%%%%%%%%%%%%%%%%%%%%%%%%%%%%%%%%%%%%%%%%%%%%%%%%%%%%%%%%
\section{Outlook}

In this paper it was demonstrated that optics of a sufficiently high quality are available for prototype-scale experiments with 25\,mm diameter optics and a beam radius of 1.9\,mm though the polarisation optics. While further testing is required, there is considerable promise of being able to scale these results for a detector such as the Einstein Telescope with 10\,km arms and a beam radius of 6\,mm at the central beam-splitter \cite{ETDesign}. An approximate increase in optical aperture size by a factor of three would be relatively straightforward for the PBS as thin film polarisers can be scaled to larger optics and QWP optics are currently available with an aperture of 96mm but with a lower birefringence specification of $\case{\lambda}{4} + \case{\lambda}{200}$ \cite{TowerOptic}. 

\section{Conclusion}

While many other speedmeter designs exist, the polarisation Sagnac remains the least technologically complex and most compatible with existing infrastructure. The main limiting factors for a polarisation Sagnac speedmeter are cavity birefringence in the interferometer arms and the imperfections of the PBS and QWP polarisation optics.

In this work we investigated, with a series of experiments and optical path modelling, how polarisation effects in a polarisation Sagnac speedmeter can couple Michelson signal fields to the interferometer readout, a process that is known to  degrade the speedmeter performance due to imperfect cancellation of quantum radiation pressure noise. It was shown that for a polarisation Sagnac speedmeter utilising the PBS characterised in this work at an optimal angle of incidence the output light would contain $<0.1\%$ of Michelson signal. This is a level of imperfection that would not critically impact the quantum noise limited sensitivity of this detector design.

We have shown that it is possible to mitigate the effects arising from arm cavity birefringence by adjusting the QWP. With such a cancellation the limitation remains solely due to the finite extinction ratio of the PBS optic. This removes a significant potential disadvantage of the polarising Sagnac scheme and hence increases the feasibility of fabricating low-birefringence cavity mirrors with the required performance.

While the availability of polarising optics with the required combination of optical quality and aperture size needs to be demonstrated there is considerable promise as for the thin film polariser this should be relatively straightforward. For the QWP some development is required to produce suitable components. Leaving the polarisation Sagnac speedmeter as a strong candidate for future gravitational wave detectors with reduced quantum noise.

\ack
This work was supported by the Science and Technology Facilities Council (Grant ref: ST/N005422/1) and through STFC Studentship award number 1653071. The authors would like to acknowledge the useful advice and guidance from S. Danilishin and S. Hild and to thank S. Leavey and J. Steinlecher for providing feedback on the manuscript. This paper has LIGO document number P2000286.

\section*{References}
\bibliographystyle{unsrt}
\bibliography{Bib}

\begin{thebibliography}{10}

\bibitem{aLIGO}
J.~Aasi {\it et al}~(The LIGO Scientific~Collaboration).
\newblock Advanced {LIGO}.
\newblock {\em Class. Quantum Grav.}, 32:074001, 2015.

\bibitem{aVirgo}
F.~Acernese {\it et al}~(Virgo~Collaboration).
\newblock {Advanced {V}irgo: {A} {S}econd-Generation Interferometric
  Gravitational Wave Detector}.
\newblock {\em Class.Quantum Grav.}, 32(2):024001, 2015.

\bibitem{O1O2cat1}
B.~P.~Abbott {\it et al}~(LIGO Scientific~Collaboration and Virgo
  Collaboration).
\newblock {GWTC-1: A Gravitational-Wave Transient Catalog of Compact Binary
  Mergers Observed by LIGO and Virgo during the First and Second Observing
  Runs}.
\newblock {\em Phys. Rev. X}, 9:031040, Sep 2019.

\bibitem{GWTC2}
R.~Abbott {\it et al}~(LIGO Scientific~Collaboration and Virgo Collaboration).
\newblock {GWTC}-2: Compact binary coalescences observed by {LIGO and Virgo}
  during the first half of the third observing run.
\newblock 10 2020.

\bibitem{InterferometryLivRev}
A.~{Freise} and K.~{Strain}.
\newblock {Interferometer Techniques for Gravitational-Wave Detection}.
\newblock {\em Living Review in Relativity}, 13(1), 2010.

\bibitem{Buikema20}
A.~Buikema {\it et al}.
\newblock Sensitivity and performance of the {Advanced LIGO} detectors in the
  third observing run.
\newblock {\em Phys. Rev. D}, 102:062003, Sep 2020.

\bibitem{Acernese19}
F.~Acernese {\it et al}.
\newblock Increasing the astrophysical reach of the {Advanced Virgo} detector
  via the application of squeezed vacuum states of light.
\newblock {\em Phys. Rev. Lett.}, 123:231108, Dec 2019.

\bibitem{Bode20}
N.~Bode {\it et al}.
\newblock Advanced {LIGO} laser systems for {O3} and future observation runs.
\newblock {\em Galaxies}, 8(4), 2020.

\bibitem{Tse19}
M.~Tse {\it et al}.
\newblock Quantum-enhanced {Advanced LIGO} detectors in the era of
  gravitational-wave astronomy.
\newblock {\em Phys. Rev. Lett.}, 123:231107, Dec 2019.

\bibitem{Mehmet20}
M.~{Mehmet} and H.~{Vahlbruch}.
\newblock The squeezed light source for the {Advanced Virgo} detector in the
  observation run {O3}.
\newblock {\em Galaxies}, 8(4), 2020.

\bibitem{Corbitt04}
T.~{Corbitt} and N.~{Mavalvala}.
\newblock {Quantum Noise in Gravitational Wave Interferometers}.
\newblock {\em Journal of Optics B}, 6(8), 2004.

\bibitem{FirstSpeedmeter}
V.~{Braginsky} and F.~{Khalili}.
\newblock {Gravitational Wave Antenna with {QND} Speed Meter}.
\newblock {\em Phy. Lett. A}, 147:251, 1990.

\bibitem{QMTLivRev}
S.~{Danilishin} and F.~{Khalili}.
\newblock {Quantum Measurement Theory in Gravitational Wave Detectors}.
\newblock {\em Living Reviews in Relativity}, 15(5), 2012.

\bibitem{ET10}
M.~Punturo {\it et al}.
\newblock {The Einstein Telescope: a third-generation gravitational wave
  observatory}.
\newblock {\em Classical and Quantum Gravity}, 27(19):194002, Sep 2010.

\bibitem{DanilishinPolSM}
S.~{Danilishin}.
\newblock {Sensitivity Limitations in Optical Speed Meter Topology of
  Gravitational-Wave Antennas}.
\newblock {\em Phy. Rev. D}, 69:102003, 2004.

\bibitem{ChenRing}
Y.~{Chen}.
\newblock {Sagnac Interferometer as a Speed-Meter-Type, Quantum-Nondemolition
  Gravitational-Wave Detector}.
\newblock {\em Phy. Rev. D}, 67:122004, 2003.

\bibitem{PurdueChenSlosh}
P.~{Purdue} and Y.~{Chen}.
\newblock {Practical Speed Meter Designs for Quantum Nondemolition
  Gravitational-Wave Interferometers}.
\newblock {\em Phy. Rev. D}, 66:122004, 2002.

\bibitem{Huttner}
S.~Huttner {\it et al}.
\newblock {Candidates for a Possible Third-Generation Gravitational Wave
  Detector: Comparison of ring-Sagnac and sloshing-Sagnac speedmeter
  interferometers}.
\newblock {\em Class. Quantum Grav.}, 34(2), 2016.

\bibitem{DanilishinPolCirc}
S.~{Danilishin}, E.~{Knyazev}, N.~{Voronchev}, F.~{Khalili}, C.~{Gr{\"a}f},
  S.~{Steinlechner}, J.~{Hennig}, and S.~{Hild}.
\newblock {A New Type of Quantum Speed Meter Interferometer Measuring Speed to
  Search for Intermediate Mass Black Holes.}
\newblock {\em Light Science and Applications}, 7(11), 2018.

\bibitem{EPR}
E.~{Knyazev}, S.~{Danilishin}, S.~{Hild}, and F.~{Khalili}.
\newblock {Speedmeter Scheme for Gravitational-Wave Detectors Based on EPR
  Quantum Entanglement}.
\newblock {\em Physics Letters A}, 382(33):2219 -- 2225, 2018.

\bibitem{AsymmerticSSM}
S.~{Danilishin}, C.~{Gräf}, S.~{Leavey}, J.~{Hennig}, E.~{Houston},
  D.~{Pascucci}, S.~{Steinlechner}, J.~{Wright}, and S.~{Hild}.
\newblock {Quantum Noise of Non-Ideal {S}agnac Speed Meter Interferometer with
  Asymmetries}.
\newblock {\em New Journal of Physics}, 17, 2015.

\bibitem{Layertech}
{Layertec}.
\newblock https://www.layertec.de/en/.

\bibitem{Calcite}
K.D. {Skeldon}, D.A. {Clubley}, G.P. {Newton}, S.~{Thieux}, M.~{Von Gradowski},
  and B.W. {Barr}.
\newblock {Measurements of an Ultra-Low Loss Polarizer for $\lambda =$1064 nm
  Using a High Finesse Optical Cavity}.
\newblock {\em Journal of Modern Optics}, 48:695--702, Mar 2001.

\bibitem{UOptic}
{Union Optic}.
\newblock https://www.u-optic.com/.

\bibitem{PolShurcliff}
W.~{Shurcliff}.
\newblock {\em {Polarized Light Production and Use}}.
\newblock Harvard University Press, 1962.

\bibitem{BirefringenceMoriwaki}
S.~{Moriwaki}, H.~{Sakaida}, T.~{Yuzawa}, and N.~{Mio}.
\newblock {Measurement of the Residual Birefringence of Interferential Mirrors
  Using {Fabry-Perot} Cavity}.
\newblock {\em Appl. Phys. B}, 65:347--350, 1997.

\bibitem{BirefringenceBrandi}
F.~{Brandi}, F.~{Della Valle}, A.M. {De Riva}, P.~{Micossi}, F.~{Perrone},
  C.~{Rizzo}, G.~{Ruoso}, and G.~{Zavattini}.
\newblock {Measurement of the Phase Anisotropy of Very High Reflectivity
  Interferential Mirrors}.
\newblock {\em Applied Physics B}, 65:351--355, Sept 1997.

\bibitem{BirefringenceJacob}
D.~{Jacob}, M.~{Vallet}, F.~{Bretenaker}, A.~{Le Floch}, and M.~{Oger}.
\newblock {Supermirror Phase Anisotropy Measurement}.
\newblock {\em Opt. Lett.}, 20(7):671--673, Apr 1995.

\bibitem{BirefringenceReview}
F.~{Bielsa}, A.~{Dupays}, M.~{Fouche}, R.~{Battesti}, C.~{Robilliard}, and
  C.~{Rizzo}.
\newblock {Birefringence of Interferential Mirrors at Normal Incidence
  (Experimental and Computational study)}.
\newblock {\em Appl. Phys. B}, 97(2):457--463, 2009.

\bibitem{Cole16}
G.~Cole {\it et al}.
\newblock {High-Performance Near- and Mid-Infrared Crystalline Coatings}.
\newblock {\em Optica}, 3:647656, 2016.

\bibitem{Winkler}
W.~{Winkler}, A.~{Rüdiger}, R.~{Schilling}, K.A. {Strain}, and K.~{Danzmann}.
\newblock {Birefringence-induced losses in interferometers}.
\newblock {\em Optics Communications}, 112(5-6):245--252, 1994.

\bibitem{CDS}
R.~{Bork}, R.~{Abbott}, D.~{Barker}, and J.~{Heefner}.
\newblock {An Overview of the {LIGO} Control and Data Acquisition Systems}.
\newblock {\em Proc. ICALEPCS}, pages 19--23, Jan 2001.

\bibitem{SpencerPhD}
A.P. {Spencer}.
\newblock {\em {Advanced techniques in laser interferometry for current and
  future gravitational wave detectors}}.
\newblock {PhD Thesis}, {University of Glasgow}, 2020.

\bibitem{ETDesign}
M.~Punturo {\it et al}.
\newblock Einstein gravitational wave telescope ({ET}) conceptual design study.
\newblock http://www.et-gw.eu/index.php/relevant-et-documents, 2011.

\bibitem{TowerOptic}
{Tower Optic}.
\newblock https://www.toweroptical.com/4-zero-order-waveplates/.

\end{thebibliography}

\end{document}